\documentclass[12pt]{iopart}
\expandafter\let\csname equation*\endcsname\relax
\expandafter\let\csname endequation*\endcsname\relax
\usepackage{iopams}  
\usepackage{graphicx}
\usepackage{makecell}
\usepackage{ulem} 
\usepackage{amsmath}

\begin{document}

\title[Early warning signals]{Early warning signals for critical transitions in complex systems}

\author{Sandip V. George$^{1,2}$, Sneha Kachhara$^{3,4}$ and G. Ambika$^{3,5}$}

\address{$^1$ Department of Psychiatry, Interdisciplinary Center Psychopathology and Emotion regulation (ICPE), University of Groningen, University Medical Center Groningen (UMCG), Groningen, The Netherlands}
\address{$^2$ Department of Computer Science, University College London, United Kingdom}

\address{$^3$ Department of Physics, Indian Institute of Science Education and Research (IISER) Tirupati, India}
\address{$^4$ Department of Molecular Biosciences, Northwestern University, Evanston, USA}
\address{$^5$ School of Physics, Indian Institute of Science Education and Research (IISER) Thiruvananthapuram, India}
\ead{g.ambika@iisertvm.ac.in}

\vspace{10pt}
\begin{indented}
\item[] 
\end{indented}

\begin{abstract}
In this topical review, we present a brief overview of the different methods and measures to detect the occurrence of critical transitions in complex systems. We start by introducing the mechanisms that trigger critical transitions, and how they relate to early warning signals (EWS) and mention briefly the conventional measures based on critical slowing down as computed from data and applied to real systems. We then present in detail the approaches for multivariate data, including those defined for complex networks. More recent techniques like the warning signals derived from the recurrence pattern underlying the data, are presented in detail as measures from recurrence plots and recurrence networks. This is followed by a discussion on how methods based on machine learning are used most recently, to detect critical transitions in real and simulated data. Towards the end, we summarise the issues faced while computing the EWS from real-world data and conclude with our outlook and perspective on future trends in this area.
 
\end{abstract}

%
%
%
%
%

\section{Introduction}
Many real-world systems like ecosystems, climate, biological systems and engineering systems are complex and are characterised by the ability to self-organise and adapt under external influences or environments. However, due to their multi-component nature, nonlinear behaviour and complex pattern of interactions, they also show multi-stability that creates possibility of sudden, unexpected shifts from one dynamical state to a different one. These abrupt transitions are referred to as \textit{critical transitions}. Many of these transitions occur with seemingly tiny changes in intrinsic or extrinsic conditions \cite{scheffer2009critical} and are difficult to anticipate. Quite often, such transitions cause natural calamities or failures in engineered systems and infrastructure. Therefore, predicting their occurrence is essential to mitigate disaster impacts and to manage risk \cite{lenton2019climate,suweis2014early}. From core collapse in massive stars \cite{fryer2004collapse}, erupting volcanoes \cite{rohmer2016anticipating} to cardiac arrests \cite{nannes2020early}, such sudden phenomena are prevalent in real-world complex systems at different scales \cite{li2004multi}. The extinction of species \cite{rietkerk2004self} and resilience loss in marine and terrestrial ecosystems \cite{rocha2022ecosystems}, glacial retreat and warming oceans\cite{boers2021critical}, onset and withdrawal of monsoon \cite{thomas2015early}, natural hazards like cyclones \cite{prettyman2018novel} and earthquakes \cite{fan2020statistical}, regression of productive farmland or fisheries \cite{litzow2016early}, surge of epidemics \cite{drake2019statistics}, sudden seizures in epilepsy \cite{jirsa2014nature}, sudden changes in mental state in psychiatry \cite{wichers2016critical, kunkels2021efficacy} and,  failures in electric power grids \cite{ren2015early}, transportation networks \cite{lacasa2009jamming} and thermoacoustic systems \cite{pavithran2021critical} are all such critical transitions that pose threats to humanity in many ways. 

These sudden transitions are often difficult to predict from the average response of the system. Hence any characteristics that can be observed and captured as indicators of the approaching transition are very relevant. More so, since it is often difficult to revert the system to the previous state once a critical transition has occurred. Such characteristics, when obtained ahead of time, are termed as Early Warning Signals (EWS) \cite{scheffer2009early}. There has been considerable progress in the search for effective EWS in aforementioned areas in the recent years, derived from statistical and nonlinear time series approaches. With the availability of abundant data and increased computational power, many data-driven approaches, especially those based on machine learning, are now being explored.

An important step in isolating effective EWS is to identify the mechanisms that cause the transitions. The critical transitions in complex dynamical systems are generally understood using the idea of tipping. The term \textit{tipping point} refers to a threshold or critical value (of a parameter) near a transition, at which the system becomes highly vulnerable and can be pushed suddenly to a different state that can be pathological, disastrous, or undesired \cite{van2016you}. Several causes can lead to these sudden changes. They can be extreme events in purely stochastic processes or can arise from external phenomena, such as the Cretaceous-Paleogene mass extinction event \cite{schulte2010chicxulub}. But when they arise from changes in the inherent dynamics of the system, the system is said to undergo \textit{tipping}. Three main types of tipping are identified in complex dynamical systems, namely bifurcation-induced, noise-induced and rate-induced tipping \cite{ashwin2012tipping, ambika2021tipping}. The origin of this classification stems from dynamical systems theory, where the evolution of a given system can be described with differential equations that admit a set of parameters depending on details of the system. A \textit{bifurcation-induced tipping} occurs when one or more of the parameters of a system reach critical values that cause a sudden shift or bifurcation to a new stable state. A \textit{noise-induced tipping} occurs when random perturbations force the system to tip from one state to another in a multi-stable system with two or more stable states co-existing for the same set of system parameters. In both of these cases, the  alternate state exists and the system can be considered to be quasi-static \cite{ashwin2012tipping}. However, in the case of \textit{rate-induced tipping}, one or more parameters change at a rate fast enough to prevent a steady evolution of the system, causing system collapse or a critical transition even when there is no pre-existing alternate state for the particular value of parameter(s) \cite{ashwin2012tipping,ritchie2016early}. It is also worth mentioning that an observed transition may also arise from a combination of these phenomena. \\

A mathematical model of a dynamical system should be sufficient to predict possible future trajectories, given the knowledge of the current state and system parameters. However, in most cases we do not have a suitable model with dynamical equations, to begin with. Moreover, real systems are generally subject to fluctuations, both intrinsic and extrinsic and therefore face the risk of critical transitions. The generic features exhibited by the systems near many such transitions can be quantified using measures that can be estimated from observational or measured data or time series of the response of the systems. These data-driven approaches are now established as an efficient way to arrive at early indicators of transitions or EWS for a range of complex systems\cite{Liu2022,dablander2022anticipating,bury2021deep}.

While sudden transitions are well studied for simple isolated systems, there exist many challenges in understanding how the heterogeneous structure in complex systems, such as networks of species, habitats, climate or society, undergo transitions in response to changing conditions and perturbations. In such connected systems, the tipping of one unit can sometimes induce cascades of tipping or domino effect. For instance, changing the population of a predator in an ecosystem can cause cascades of shifts across complex food webs \cite{carpenter2008leading}. The EWS for such transitions often involve the relationship between multiple variables, or the changes in the topological features of the structure of the system.

Recent research in ecology and climate has lead to the understanding of critical transitions and system resilience when subjected to changing conditions. Ecological systems such as shallow lakes, coral reefs, range-lands, woodlands, grasslands, rain-forests, salt marshes, arid and semi-arid ecosystems \cite{scheffer2009critical}, the Amazon rainforest cover\cite{boulton2022pronounced},  Greenland and West Antarctic ice sheets \cite{boers2021critical,rosier2021tipping}, changes in the patterns of El Ni\~{n}o oscillations \cite{ludescher2014very}, Atlantic Overturning Circulation \cite{boers2021observation}, and Coral reefs \cite{lenton2008tipping} are some notable examples.  Also, in sociology, finance, psychology and neuroscience, some applications of resilience indicators and associated concepts is reported\cite{Liu2022}.

In this review we present a summary of the various methods used for predicting critical transitions. We start with a brief discussion of conventional measures reported as effective EWS in various areas of research, including measures for spatial data (Section \ref{sec:conventionalews}). We then proceed to discuss novel techniques that are developed for cases when data are multivariate because of an underlying network structure or because multiple system responses are measured (Section \ref{sec:multivar}). We discuss in detail the recently studied measures based on the recurrence patterns in the
data using recurrence plots and recurrence networks (Section 4). These provide insights
into the recurrence of the states reconstructed through delay embedding. We then present
the detection of critical transitions using machine learning techniques (Section 5), that are
relatively less explored but are useful in detecting transitions in different contexts.
Moreover they are mostly model independent and hence capable of predicting transitions
that are not necessarily preceded by CSD. We also caution about potential pitfalls in detecting critical transitions using EWS (Section \ref{sec:reliability}). Towards the end we discuss recent trends and possible avenues for future research in extending the applicability of EWS and enhancing our understanding of transitions in complex systems (Section \ref{sec:dis}).

\section{Conventional Early Warning Signals}
\label{sec:conventionalews}
It is well established that near some bifurcation-induced transitions, dynamical systems exhibit the phenomena of ``critical slowing down" (CSD). This is consequent to an increase in the relaxation time seen close to bifurcation points in dynamical systems whereby once perturbed, the system takes long time to relax \cite{scheffer2009critical}.  A list of bifurcations preceded by CSD along with the nature of these transitions is listed in detail in \cite{thompson2011predicting}. Though a large number of bifurcations exhibit CSD, EWS literature is mostly centred around fold bifurcations\cite{scheffer2009critical}. Close to the bifurcation point, a very small change in parameter or a small perturbation can cause a large shift or transition to another stable state. Hence fold bifurcations illustrate the mechanism of sudden transitions very well, exhibit CSD and can model a large variety of transitions across fields \cite{kuehn2011mathematical}. For this reason it remains a model of choice, despite some criticism in the past regarding its overuse \cite{zahler1977claims}.

The phenomenon of CSD forms the basis for many measures useful as EWS, including increased autocorrelation and variance \cite{scheffer2009early,ditlevsen2010tipping}. The autocorrelation measures the linear correlation between two points that are separated by a temporal lag in a time series. An increased relaxation time would result in a larger correlation between nearby points in the time series, and is true in particular for correlations of adjacent points, i.e. the autocorrelation at lag-1 (ACF(1)). Thus an increasing trend in ACF(1) serves as EWS for such transitions. Since the impact of perturbations takes longer to decay as the system approaches a critical transition, the variance of the system’s response also increases \cite{scheffer2009early,dakos2012robustness}. The changes in ACF(1) and variance are the most popularly studied EWS, and are successfully applied to detect critical transitions in a wide range of fields \cite{wichers2016critical,maturana2020critical, george2020early,boers2018early, harris2020early}. The increasing trends in ACF(1) and variance as EWS can be formally derived from the theory of stochastic dynamical systems\cite{ditlevsen2010tipping, qin2018early}. Starting from the linearised dynamics about the stable point of a system perturbed with white noise,  it is shown that the variance approaches $\infty$ and the ACF approaches 1 as the system approaches certain types of bifurcations \cite{bury2020detecting, boettner2022critical}.

Apart from the variance, higher order moments of the amplitude distribution of the system response, such as the skewness and kurtosis, change as the system approaches a critical point \cite{guttal2008changing}. The estimation of these quantifiers are listed in Table \ref{table:error}. These measures are computed from data using a sliding window approach, where the quantifier of interest is estimated within a window of $W$ points in the time series which is slid forward in time with a suitable time step to get the variations over time. If the quantifier shows significant trends over time, often confirmed using rank correlation coefficients like the Kendall's correlation coefficient \cite{chen2022practical}, it is taken as evidence of CSD. While this technique is widely used, several caveats exist, which are discussed in detail in Section \ref{sec:reliability}.  

The transitions that are accompanied by CSD can also be detected as changes in spectral measures. When the system exhibits CSD, the lower frequencies of the power spectrum show an increased power (called reddening) \cite{kleinen2003potential, tan2014critical}. This increase is equivalent to the increase discussed above in the lower lags of the autocorrelation, which is the Fourier transform of the power spectrum. This reddening may be quantified through an increase in the power-law scaling exponent, $\beta$, of the spectrum ($S(f) = f^-\beta$), and is used as a warning signal in a number of applications \cite{biggs2009turning,prettyman2018novel}. 

Another closely related measure is the Hurst exponent, $H$, which measures the long term fluctuations in the time series and consequently increases when the system exhibits CSD \cite{kantelhardt2001detecting}. The most popular method to determine the Hurst exponent is the detrended fluctuation analysis (DFA) \cite{bryce2012revisiting}. We note that even certain transitions, such as rate-induced tipping that are not known to be preceded by CSD, may also exhibit significant trends in many of the quantifiers mentioned above \cite{ritchie2016early}. 

With growing confidence in EWS and simultaneous rise in data availability and computational power, their applications have expanded to more areas. A highly relevant example is the abrupt climate change and its implications. Many observation-based studies have identified trends in past climate shift events \cite{brovkin2021past, wang2020early} and project dire consequences if we follow present trends \cite{wunderling2020motifs}. 

\begin{figure}
        \centering
        \includegraphics[width=0.5\textwidth]{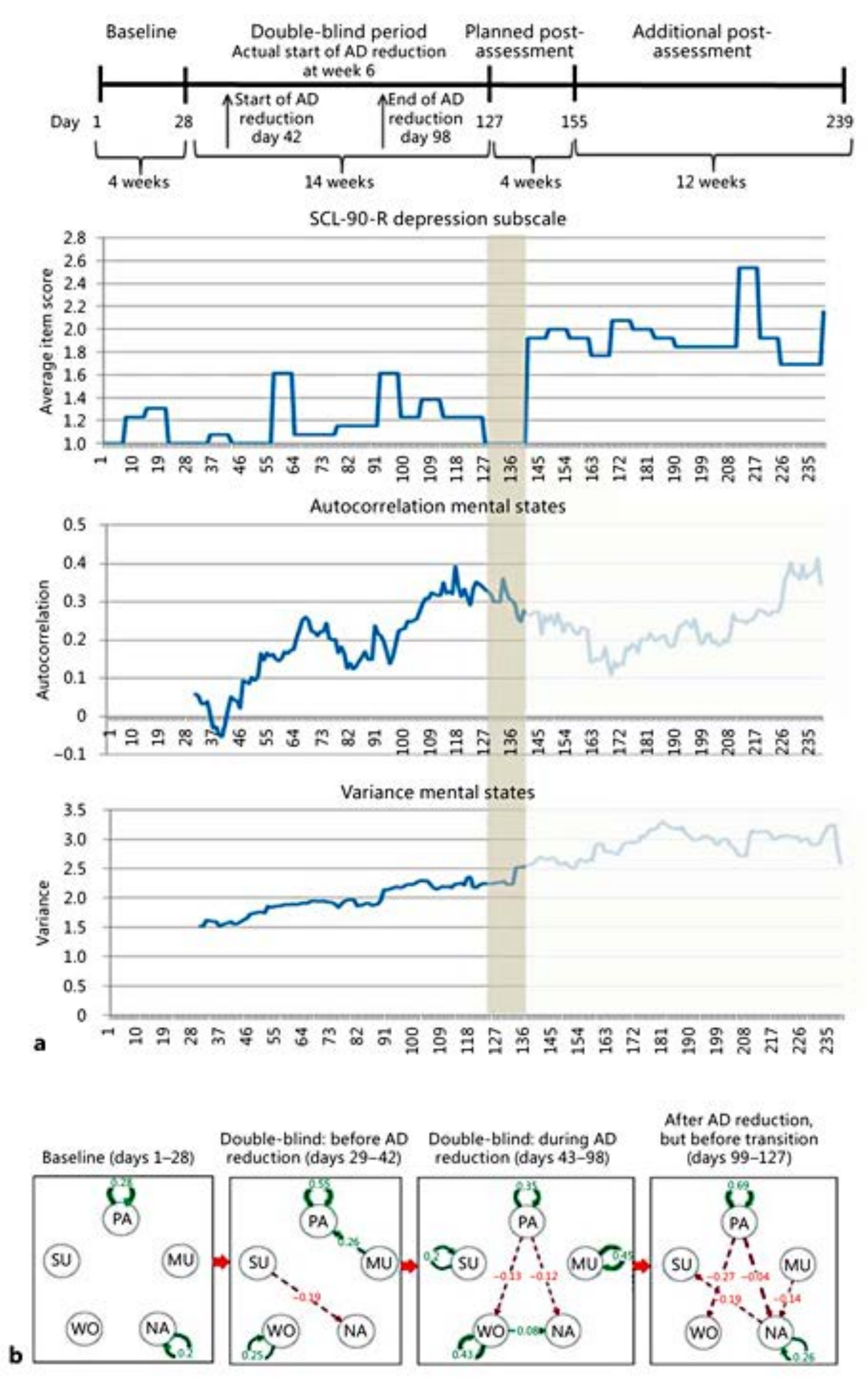}
        \caption{EWS of a critical transition towards depression, in the weeks leading up to the episode. The ACF(1) and variance calculated from momentary mood states increase in the weeks prior to the episode. The onset of the episode is shown as a grey rectangle. The bottom panels show the correlations between the different mood states at various stages of mood shift. (Credit : Wichers M, Groot P, C:, Psychother Psychosom 2016;85:114-116. Reproduced with permission from S. Karger AG, Basel \cite{wichers2016critical}).}
        \label{fig:depression}
    \end{figure}

In the health sciences, research on CSD is particularly emphasised in mental health research and seizure prediction \cite{van2014critical,maturana2020critical, bayani2017critical}. CSD based warning signals are discovered in self reported momentary mood states in depression\cite{wichers2016critical,curtiss2021rising} (See Figure \ref{fig:depression}) and bipolar disorder \cite{bos2022anticipating}. In seizure prediction, EWS are shown in computation and in-vitro studies \cite{freestone2017forward, meisel2012scaling, negahbani2015noise, jirsa2014nature, chang2018loss}, though its detection in real data remains disputed \cite{milanowski2016seizures, wilkat2019no, maturana2020critical}. EWS also reportedly precedes transitions in chronic illnesses such as asthma, cardiac arrhythmias \cite{olde2016slowing} and multiple sclerosis \cite{twose2020early}, as well as in ventricular fibrillation \cite{nannes2020early}. At a community level, EWS are shown to precede significant increases in hospital admissions for cardio-pulmonary diseases \cite{wang2018early}.\\

\begin{table}
\caption{Some of the quantifiers commonly used as EWS, their estimation and the corresponding standard error in estimation from $N$ points. The formulae used for standard errors have the underlying assumption of normality in the sampling distribution, which leads to the errors being underestimated\cite{wright2011problematic}. These values should hence serve as a lower bound on the estimated error.}             
\label{table:error}     
\centering                          
\begin{tabular}{c c c c}        
\hline\hline  
\ms
Quantifier  &Estimation &Error & Reference   \\ [0.5ex] 
\hline 
\ms
Autocorrelation & $\frac{1}{N}\sum_{i=1}^{N}\frac{ (X_i-\mu)(X_{i+\tau}-\mu)}{(X_i-\mu)^2}$ &$\sqrt{\frac{1-A(\tau)^2}{N-3}}$ & \cite{bonett2000sample}  \\ 
\hline
\ms 
Variance  &  $\frac{1}{N-1}\Sigma_i (X_i-\mu)^2$ &$\sqrt{\frac{2\sigma^4}{N-1}}$ & \cite{ahn2003standard}\\ 
\hline
\ms
Skewness  &$\frac{\frac{1}{N}\Sigma_{i=1}^{N}(x_i-\mu)^3}{[\frac{1}{N}\Sigma_{1}^{N}(x_i-\mu)^2]^{\frac{3}{2}}}$ &$\sqrt{\frac{6(N-2)}{(N+1)(N+3)}}$ & \cite{wright2011problematic}\\
\hline
\ms
Kurtosis  &$\frac{\frac{1}{N}\Sigma_{i=1}^{N}(x_i-\mu)^2}{[\frac{1}{N}\Sigma_{i=1}^{N}(x_i-\mu)^2]^2}-3$ &$\sqrt{\frac{24N(N-2)(N-3)}{(N+1)^2(N+3)(N+5)}}$ & \cite{wright2011problematic}\\

\hline                                   
\end{tabular}
\end{table}

Challenges in data collection and accuracy, as well as complexity of interactions in the aforementioned systems has prompted inquiry into spatial early warning signals (SEWS) \cite{kefi2014early}. Especially in ecology where lack of long temporally-resolved data is a major limitation, well resolved spatial data can be acquired for many large ecosystems. From this data (computed from snapshots of multiple time series or data spread over a large number of spatial points), SEWS can be calculated analogous to their temporal counterparts. Extensive studies on impending regime shifts in ecosystems based on simulated as well as real grid data collected with remote sensing are conducted as well\cite{nijp2019spatial}. Other examples include changes in patch-size distribution prior to desertification \cite{bird2007multi}, increasing spatial variance in conjunction with changing spatial skewness prior to vegetation collapse \cite{guttal2009spatial}, increasing spatial correlation in spatially heterogeneous systems \cite{dakos2010spatial}, increasing recovery length for populations of yeast \cite{dai2013slower} and collapsing marine benthic ecosystem \cite{rindi2017direct}, and increase in several CSD-based indicators prior to grassland-to-woodland transitions in Savanna ecosystem\cite{eby2017alternative}. 

\begin{figure}
    \centering
    \includegraphics[width=0.7\textwidth]{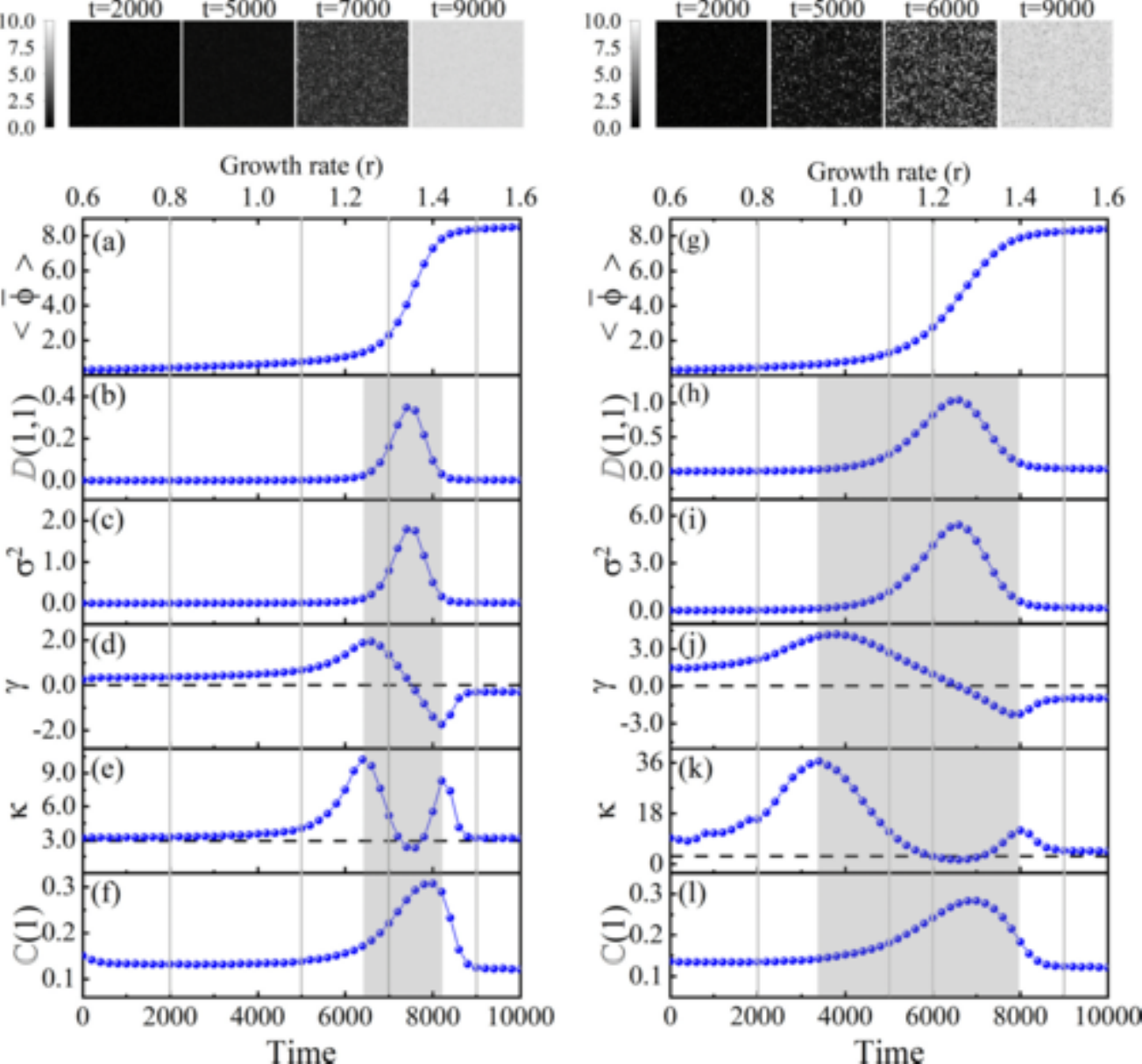}
    \caption{[(a), (g)] Transient-state ensemble average of tumour nuclei density as a function of growth rate $r$ and simulation time. The vertical gray lines correspond to the increasing growth rate $(r)$ over simulation time, where spatial snapshots are shown at the top. The snapshots are drawn in grayscale, where brighter regions represent higher tumour nucleus density. [(b)–(f), (h)–(l)] SEWS for simulated spatial snapshots data obtained from the stochastic model of Eq. (5) in \cite{ma2022spatiotemporal}. Near-neighbour spatiotemporal diffusion coefficient at lag-1 [(b), (h)], spatial variance [(c), (i)], spatial skewness [(d), (j)], spatial kurtosis [(e), (k)], and near-neighbour spatial correlation [(f), (l)] as functions of growth rate $r$ and simulation time. (Reproduced from Ma et al., 2022 \cite{ma2022spatiotemporal})}
    \label{fig:spatial_indicators_tumor}
\end{figure}

In the context of climate, cross-correlation in the finely sampled time series from different spatial locations anticipated El Ni\~{n}o oscillations up to a year in advance with 3-in-4 likelihood \cite{ludescher2014very}. On a smaller scale, Ma et al. recently demonstrated that the tumour-immune system can be treated as a two-dimensional spatially extended system that exhibits bistablity (healthy and disease state). In this context, rising spatiotemporal diffusion coefficient obtained from the spatial snapshot data is an effective indicator (see Figure \ref{fig:spatial_indicators_tumor})\cite{ma2022spatiotemporal}.

There are transitions that are not preceded by CSD \cite{boerlijst2013catastrophic,boettiger2013early} and temporal and spatial EWS may fail in these situations. Moreover, spatial indicators show different trends for different systems \cite{scheffer2009early}. Some of these limitations are addressed in later studies, such as the study of EWS prior to regime shifts that do not exhibit CSD \cite{kefi2013early} and using multivariate data to test for false alarms \cite{rohmer2016anticipating,streeter2013anticipating}. Other limitations such as the effect of large fluctuations and heterogeneous stressors \cite{genin2018spatially} cannot be avoided, which necessitates extra care with the interpretation of SEWS in such cases. An R-based toolkit for SEWS, developed by G\'{e}nin et al. \cite{genin2018monitoring} can be found at https://github.com/spatial-ews/spatialwarnings.

\section{EWS for multivariate data and complex networks}
\label{sec:multivar}
In many real-world systems, more than one response is often recorded over time and, the single variable temporal measures mentioned above are conducted independently on multiple variables\cite{bos2022anticipating}. The reliability of an EWS detection can then be estimated through the strength of EWS across multiple variables. However, a more general approach that encompasses the relationships between multivariate data is likely to be more robust. Moreover, many single variable techniques described above, such as the ACF, are developed for evenly sampled data, i.e. the time difference between adjacent observations is equal throughout the time series. In cases where multiple variables are measured at the same time instant, data gaps often occur simultaneously in all variables. In such cases, cross time series measures such as the correlation between the time series can still be reliably calculated. 

Multivariate data, in general, need not be spatial (or spatially restrictive), but can be of the different responses of the same system or of the same variable measured from different parts of an interacting network of systems. Often in real systems, multivariate data sets are available such as multiwavelength data in astronomy, ECG recordings from multiple leads, data from various spatial locations in ecology, etc.  In this section, we describe some notable examples of this approach in the wider scope of detection of critical transitions.

Many complex systems can be represented mathematically as higher-dimensional systems with random environmental perturbations,  modelled by first-order stochastic differential equations. The solution of such an equation is the probability density function $p(z, t)$ of deviations (in the state variables), $z$, from the equilibrium. As outlined in Chen et al. \cite{chen2019eigenvalues}, this function can be approximated as the solution to a linear Fokker-Planck equation. If the fluctuations have reached a stationary  distribution, the covariance matrix of the solution which gives the correlation between the state variables can be directly used to interpret the nature of dynamics. Even when the analytical expression of the covariance matrix is generally not available for a real system, we can estimate the elements of the covariance matrix, $C_{i,j}$ directly from the multivariate time series data $X^k(t)$ as the covariance between the $i^{th}$ and $j^{th}$ time series \cite{press2007numerical,park2018fundamentals}.
\begin{equation}
    C_{i,j}=Cov(X_i,X_j)=E[(X_i-E[X_i])(X_j-E[X_j])]
\end{equation}
, where E represents the expected or mean value. The eigenvalues of the covariance matrix calculated using a moving window on data can capture the CSD in the system near transitions. As shown by Chen et al., the largest eigenvalue of the covariance matrix is a useful metric to predict critical transitions because the dynamics along the dominant eigenvector becomes slower and the variance along that direction increases as the system undergoes CSD\cite{chen2019eigenvalues}. 

\begin{figure}
    \centering
    \includegraphics[width=0.7\textwidth]{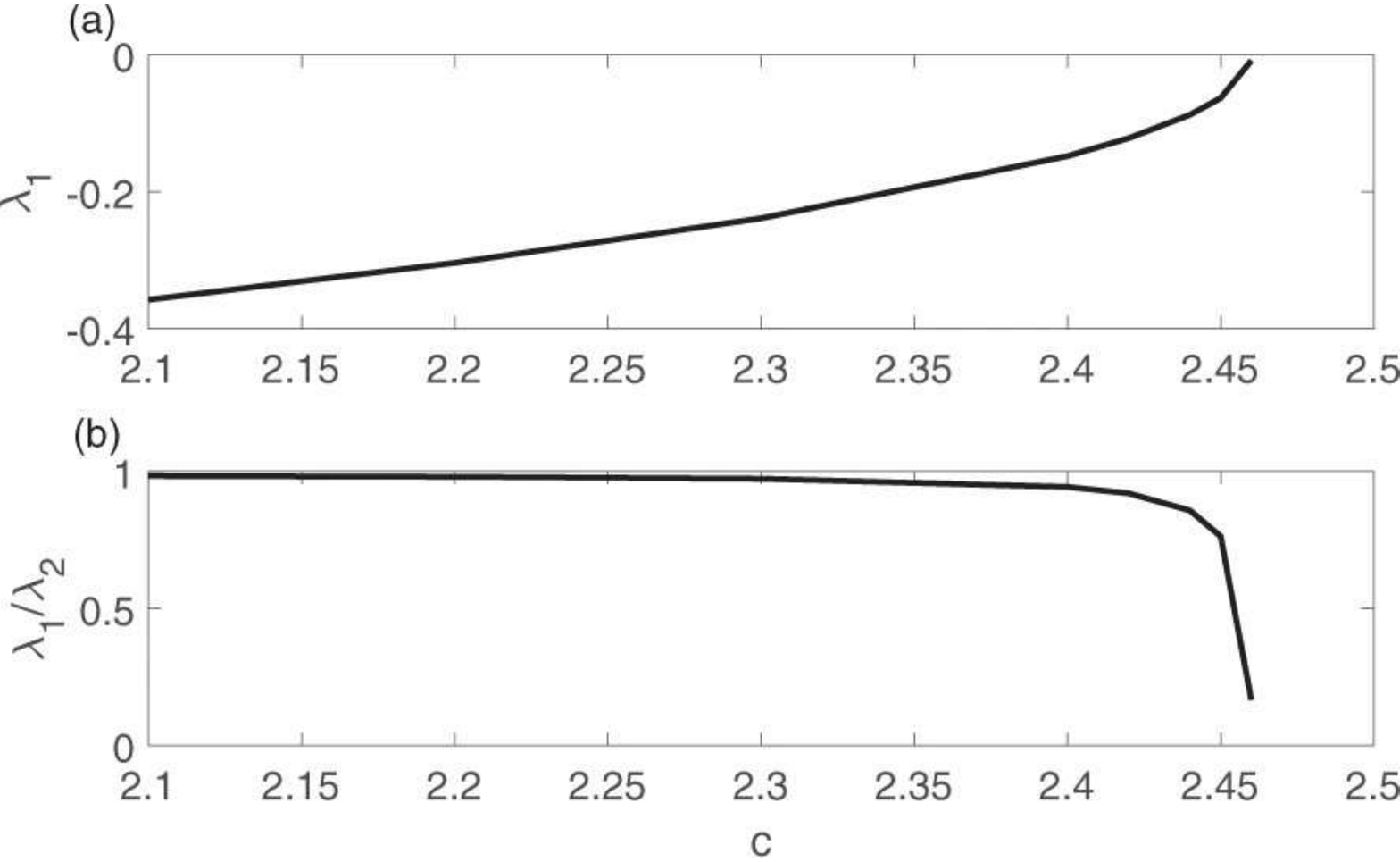}
    \caption{Variation of the (a)largest eigenvalue and (b)ratio of largest to second largest eigenvalue of the covariance matrix with varying bifurcation parameter, c (Reproduced from \cite{chen2019eigenvalues}).}
    \label{fig:eigenvalue}
\end{figure}

This is shown in Figure \ref{fig:eigenvalue} for a harvesting model, where the variation of the largest eigenvalue ($\lambda_1$) and the largest eigenvalue relative to the second largest eigenvalue ($\frac{\lambda_1}{\lambda_2}$) are shown by varying the bifurcation parameter, c. In this context, a data-driven algorithm, known as the eigensystem realisation algorithm, is useful to approximate the eigenvectors, and get EWS for the transitions \cite{ghadami2020data}. 

In addition to the maximum eigenvalue of the covariance matrix, the autocorrelation of the projection of the data on the first principal component, called degenerate fingerprinting is proposed as EWS useful for multivariate data \cite{weinans2021evaluating, suweis2014early, held2004detection}. So also, measures based on Min/Max Autocorrelation Factor (MAF) analysis and multivariate extensions of mutual information are also proposed as EWS prior to critical transitions in specific cases \cite{weinans2021evaluating, quax2013information, quax2013diminishing, marinazzo2019synergy}. 

An important class of multivariate measures rely on the framework of complex networks. In most cases, the subsystems are treated as nodes and their interaction patterns are captured as links in the network. Then the critical transitions occur as changes between emergent states, reflected as changes in network structure and dynamics \cite{Liu2022}. This framework benefits immensely from the extensive literature on graph theory that evolved systematically over the past few centuries, and the relatively recent enquiries in social networks that have led to new classes of networks and their influence on emergent dynamics\cite{newman2018networks}.

A convenient way to represent these networks is in the form of an adjacency matrix that captures links in the network as elements of the matrix, given as:

\begin{equation}
   A_{ij}=
    \begin{cases}
    1 & \quad \textnormal{nodes } i \textnormal{ and } j \textnormal{ are connected}\\
    0 & \quad  \textnormal{otherwise}
    \end{cases}
\end{equation}
As the interactions in the system change in time, so do the associated links in the network. One of the easiest way to track these changes is through the degree distribution which measures number of links per node. An interesting example is loss of low degree nodes before cascading failures in artificial networks \cite{loppini2019critical}. Another example is change in degree distribution before Atlantic Meridional Overturning Circulation (MOC) collapse \cite{van2013interaction}. Other measures such as the link density, clustering coefficient, characteristic path length, etc. can also be used to detect transitions (see Table \ref{table:network_measures}).

Very often, networks are constructed from time series at different spatial locations using lag-0 Pearson correlation coefficient so that spatio-temporal correlations are reflected in the topological properties of the network. For example, Rodr\'{i}guez et al. have shown the percolation transition in functional networks constructed from correlation of time series of nodes, to precede a bifurcation \cite{rodriguez2016percolation}. Early warning indicators of vegetation transitions \cite{yin2016network,tirabassi2014interaction} were also proposed in a similar manner. Jentsch et al. \cite{jentsch2018spatial} showed spatial correlations increasing among different nodes of a multiplex disease-behaviour network prior to a regime shift in vaccinating behaviour. 

There are instances where an increase in network connectivity, which relates to increased cross-correlation, happens before a critical transition \cite{wichers2016critical, van2014critical, tirabassi2014interaction}. In socio-ecological networks, an interaction network or community network takes care of interactions among subsystems, that represents the connectivity and strength of interactions \cite{tirabassi2014interaction, holme2023networks}. Another network-based approach is proposed by Goswami et al. \cite{goswami2018abrupt}, applied successfully to global stock indices for the detection of transitions into well-known periods of politico-economic volatility. Yet another approach based on correlation graphs is described in \cite{gorban2010correlations} and model-based early warning systems for emerging markets in \cite{ponomarenko2013early}, etc.\\

\begin{table}
    \caption{Complex Network measures useful as EWS. $A_{ij}$ represents element of the adjacency matrix for the node pair $i$ and $j$, and N is the number of nodes in the network. For a undirected, unweighted network, $A_{ij}$ = 1 if there is a link between $i$ and $j$, and 0 otherwise \cite{newman2018networks}.}             
    \label{table:network_measures}      
    \centering                          
    \begin{tabular}{c c c c}        
    \hline\hline                 
    Quantifier  &Calculation &Reference \\ [0.5ex] 
    \hline 
    Degree of node $i(k_{i})$ & $\sum_{j=1}^{N}A_{ij}$ & \cite{loppini2019critical}\\ 
    \ms
    Link Density (LD) & $(\frac{1}{N^{2}})\sum_{i,j=1}^{N}A_{ij}$ & \cite{yang2022critical}\\
    \ms
    Clustering Coefficient (CC) & $\frac{\sum_{i=1}^{N} C_{i}}{N}$, \\
     &  $C_{i}=\frac{\sum_{j,q}A_{ij}A_{jq}A_{qi}}{k_{i}(k_{i}-1)}$ & \cite{van2013interaction}\\
     \ms
    Characteristic Path Length (CPL) & $\frac{1}{N(N-1)}\sum_{i\neq j}l_{ij}$, \\
    \ms
     &  $l_{ij}$ = shortest distance between $i$ and $j$ & \cite{godavarthi2017recurrence}\\
     \ms
    Transitivity (T) & $T = \frac{\sum_{i,j,k=1}^{N}A_{j,k}A_{i,j}A_{i,k}}{\sum_{i,j,k=1}^{N}A_{i,j}A_{i,k}}$ & \cite{marwan2015complex}\\
    \hline                                   
    \end{tabular}
    \end{table}

\begin{figure}
    \centering
    \includegraphics[width=0.7\textwidth]{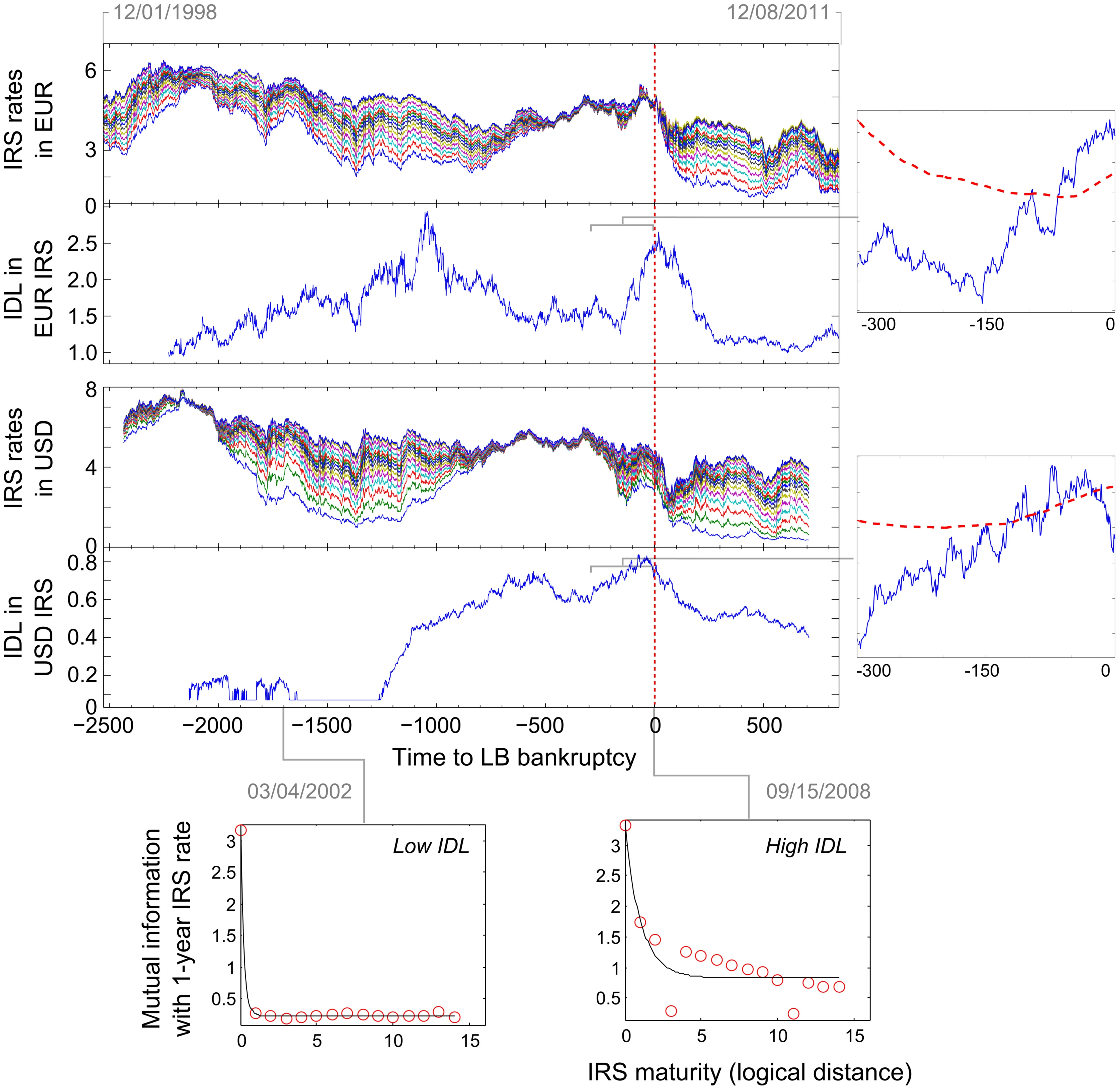}
    \caption{The upper panels show the time series of IRS rates for various maturities and the corresponding information dissipation length (IDL) for the EUR and USD markets. The inset shows the IDL and a warning threshold in the period leading up to the Lehman Brothers bankruptcy. The bottom panels show the mutual information between the rates of the 1-year maturity IRS and other maturities on two trade days. The IDL of the IRS rates across maturities for a specific trade day is estimated by fitting an exponential decay function. (Reproduced with permission from \cite{quax2013information}.)}
    \label{fig:idl}
\end{figure}

Sometimes the conventional network measures listed in table \ref{table:network_measures} are not sufficient and a closer inspection of network dynamics in terms of other properties is required. For example, a characteristic of self-organised critical transitions is long-range correlations that can be measured in terms of the transmission of information among units. Specifically, the impact of a units state on the state of the whole system at a time can be measured by their mutual information. For two random variables $X$ and $Y$, the mutual information is given by \cite{kraskov2004estimating}
\begin{equation}
    M_{X,Y}=\sum_{i,j} p_{X,Y}(i,j) log(\frac{p_{X,Y}(i,j)}{p_X(i)p_Y(j)})
\end{equation}, where $p_{X,Y}(i,j)$,  $p_X(i)$ and $p_Y(j)$ are the joint and marginal probability distributions. The characteristic distance of the decay of mutual information in the system is found by fitting an exponential decay term to the mutual information. The increase in this distance, called information dissipation length (IDL), can detect the onset of long-range correlations in the system that precede critical transitions. This is applied to measure the IDL of risk-trading among banks by calculating the IDL of the returns of interest-rate swaps (IRS) across maturities, for the EUR and USD markets using data of the daily prices of IRSs in the USD and EUR currency for maturities over 12 years \cite{quax2013information} (See Figure \ref{fig:idl}).

It is proposed that in networked systems where little information is available about the structure, EWS for the collapse of networked systems can be obtained from  their structural instability. The existence of cycles in the interaction network is one property that adds stability to the system since it implies self-sustaining structures. Thus, for a networked dynamical system the transition to instability is concurrent with that from a system with cycles to one without cycles. The corresponding EWS will be that which detects the last surviving cycle in the network. This is illustrated using the model of co-evolutionary ecosystems and applied to systems of species evolution, epidemiology, and population dynamics \cite{horstmeyer2020predicting}.

Modelling systemic risk in terms of network structure is another popular approach, especially in the context of economics and finance \cite{caccioli2018network, squartini2013early}. Wunderling et al. identified the presence of motifs (frequently repeated small structures in the network) as a cause for decreased robustness \cite{wunderling2020motifs}. Sometimes the transition depends crucially on the dynamics of a critical node. For example, reliable sensor species in ecological systems is preferred for testing an upcoming transition in relatively small natural networks \cite{ghadami2020data}. In these networks, conventional EWS can be used once the most reliable species is identified. Another example of a critical component is in chemical networks where recovery times act as EWS \cite{maguire2020early}. We note that in both cases, the actual transition is brought about by external, environmental influence, and hence can be considered to be instances of noise-induced tipping. 

 \section{Measures based on recurrences}
 \label{sec:rqa}
The conventional EWS are computed from the time series i.e. the sequence of observed or measured data. Most of these identify patterns in statistical measures over time in the (often one-dimensional) time series. A major drawback then arises from the underlying assumptions of linearity in the quantifiers used, which can cause errors in the derived conclusions. An alternative approach, that does not make these assumptions, is rooted in the recurrence of states in dynamical systems theory. This method of nonlinear time series analysis works by recreating the overall multidimensional dynamics of the system from one-dimensional data.

A dynamical system traces out trajectories in state space as it evolves in time. When these trajectories are bounded, they form a geometrical structure in the state space called attractor, which is populated by states of the system at different times. When the system is in a periodic state, the same trajectories are repeated in time. When the system exhibits more complex dynamical states such as quasiperiodic or chaotic behavior, the attractor exhibits finer structures. However, since the system is bounded it eventually revisits the same region of state space over time, as guaranteed by the Poincar\'{e} recurrence theorem \cite{marwan2007recurrence}. Even in the presence of small stochastic influences, the approximate structure mostly remains tractable with enough data. If the associated structural properties of the attractor change as the system approaches a transition, these are reflected in the recurrence patterns in the data. This useful insight can be exploited by visualising the patterns of recurrences and quantifying them using the framework of recurrence plots (RP) and recurrence networks (RN) \cite{pavithran2021critical}.\\ 

\subsection{Measures from Recurrence plots}

For most nonlinear time series analysis approaches, the one-dimensional data is first embedded into higher-dimensional space using the method of delay embedding \cite{ambika2020methods}. The Taken's delay embedding theorem guarantees that the dynamics resulting from such an embedding is topologically equivalent to the dynamics of the full multidimensional system \cite{ambika2020methods}. In order to study the recurrences in the system, these embedded M-dimensional vectors are used to construct the recurrence matrix $R$. The elements of this matrix are defined as
\begin{equation}
    R_{ij} = \Theta \left ( \varepsilon-\left \| \vec{v_{i}}- \vec{v_{j}} \right \| \right )
\label{eq:rec_matrix}
\end{equation}
Here $\vec{v_{i}}$ and $\vec{v_{j}}$ are the corresponding vectors for two points $i$ and $j$ in the reconstructed state space, $\Theta$ is the Heaviside step function, and $\epsilon$ is a threshold chosen to define when a state recurs \cite{marwan2007recurrence}. A representation of $R$, known as Recurrence Plot (RP), is a 2-dimensional discrete realisation that gives a visual representation of the recurrences, with the 1s in $R$ taken as black points and 0s as white spaces. Then the geometrical patterns in RP can be quantified using measures such as Recurrence Rate (RR), Determinism (DET), Laminarity (LAM), and Entropy (ENT) as indicated in Table \ref{table:recurrence_measures} \cite{marwan2007recurrence}.

\begin{table}
    \caption{Recurrence based quantifiers useful as EWS. $R_{ij}$ are the elements of the recurrence matrix, as defined in eq. (\ref{eq:rec_matrix}). $P(l)$ represents distribution of diagonal lines and $P(v)$ represents that of vertical lines in the RP. For details, see \cite{marwan2007recurrence}.}         

    
    \label{table:recurrence_measures}      
    \centering                          
    \begin{tabular}{c c c}        
    \hline\hline                 
    Quantifier  &Calculation &References   \\ [0.5ex] 
    \hline 
    Recurrence Rate (RR) & $\frac{1}{N^2}\sum_{i,j = 1}^{N}R_{ij}$ & \cite{pavithran2021critical}\\
    \hline
    \ms
    Determinism (DET) & $\frac{\sum_{l=lmin}^{N}lP(l)}{\sum_{ij}R_{ij}}$ & \cite{marwan2013recurrence,george2020early,westerhold2020astronomically}\\
    \hline
    \ms
    Laminarity (LAM) & $\frac{\sum_{v=vmin}^{N}vP(v)}{\sum_{v=1}^{N}vP(v)}$ & \cite{marwan2013recurrence,george2020early}\\
    \hline
    \ms
    Entropy (ENT) & $-\sum_{l=l_{min}}^{N}P(l)\ln P(l)$ & \cite{savari2016early}\\
    \hline                                   
    \end{tabular}
\end{table}

\begin{figure}
    \centering
    \includegraphics[width=\textwidth]{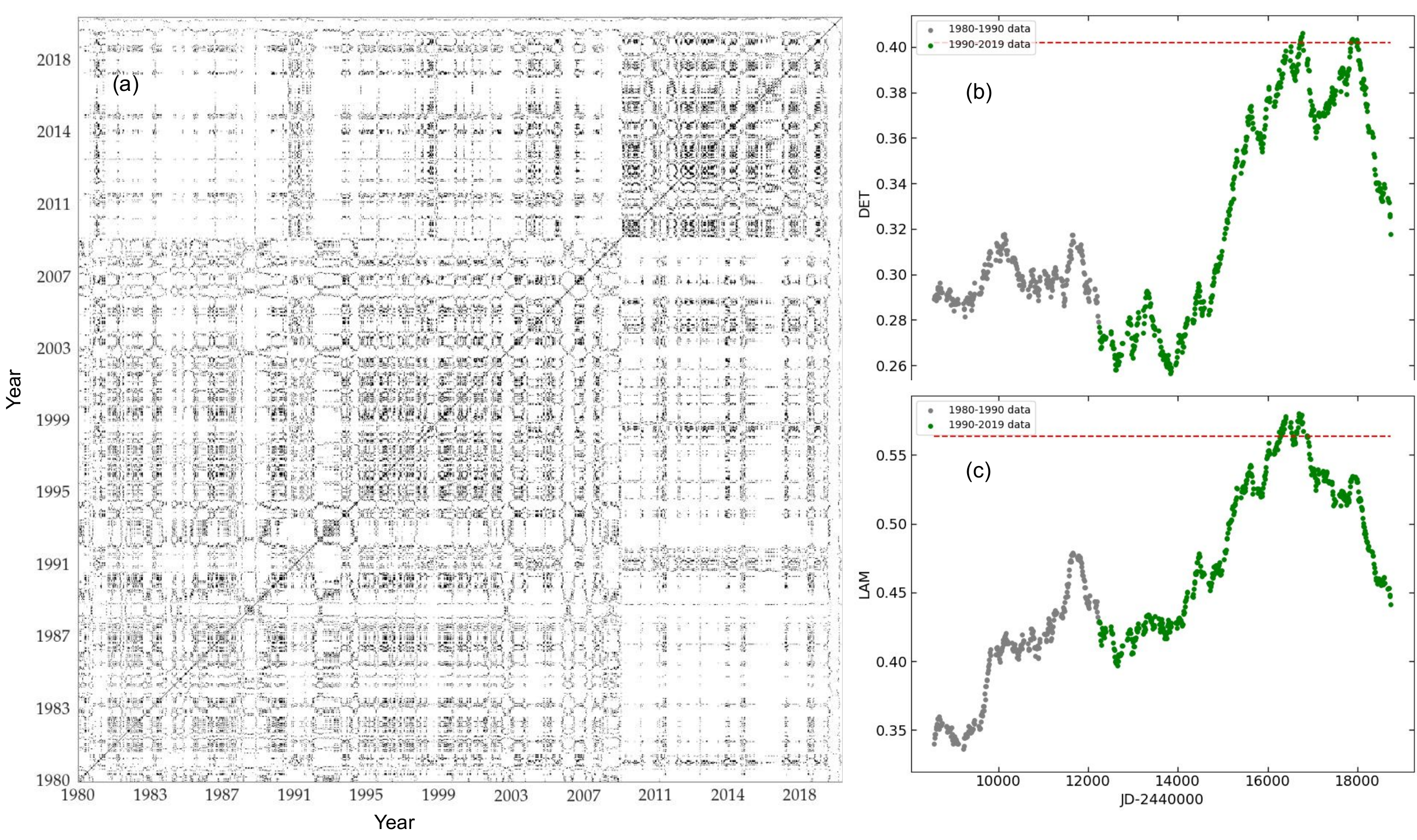}
    \caption{(a) recurrence plot for the variable star Betelgeuse constructed from its brightness data (V-band) from 1980 to 2019, prior to a transition resulting in a major dimming in the star. (b) DET and (c) LAM measures calculated from the RP. Both DET and LAM show definite trends over time, and reach the 95\% confidence interval shown in red. (Credit : George SV, Kachhara S, Misra R \& Ambika G, A\&A, 640 L21 2020 Reproduced with permission $\copyright$ ESO\cite{george2020early}).}
    \label{fig:betelgeuse}
\end{figure}

A typical RP for the variable star Betelgeuse is shown in Figure \ref{fig:betelgeuse} (a). In general, the patterns formed by recurrence of points can provide evidence for determinism or stochasticity in the dynamics of the system. The diagonal lines in RP indicate that the trajectory visit the same region of the state space at distinct times and is quantified as DET. The presence of vertical or horizontal lines, measured by LAM, means the system persists in the same region for some time. On the other hand, isolated points may occur under noisy fluctuations, where DET indicates statistical correlations in the data. Since the recurrence-based measures are directly linked to the nature of dynamics in state space, any change in the trajectory in this space manifests as change in patterns of the RP, and consequently as variations in the measures computed from it. For the variable star Betelgeuse, the recurrence measures DET and LAM act as an EWS of a possible critical transition \cite{george2020early} (See Figure \ref{fig:betelgeuse} (b) and (c)).

To estimate the changes in these RP measures over time, a sliding window analysis of the time series near transition is carried out \cite{marwan2013recurrence}. The success of recurrence based measures depends on careful selection of embedding parameters and threshold $\varepsilon$. A detailed discussion on this and established practices can be found in \cite{marwan2011avoid}.  The applications of these measures include detection of past climate states \cite{westerhold2020astronomically}, detecting regime shifts in thermoacoustic systems \cite{pavithran2021critical} and agglomeration of particles in a chemical system \cite{savari2016early}. 

More recently, another measure called Lacunarity ($\Lambda$) was used to detect regime shifts in complex systems, especially to predict the onset of thermoacoustic instability \cite{braun2021detection}. It is a multi-scale recurrence quantifier that quantifies the heterogeneity of recurrent temporal patterns representing different segments of the trajectory in the embedded state space and their self-similarity.
\subsection{Measures from Recurrence networks}

A further important extension in this direction is that of a recurrence network (RN), which is a complex network that can be constructed from RP with its adjacency matrix $A_{ij}$ derived from $R_{ij}$ as \cite{donner2010} 
\begin{equation}
A_{ij} = R_{ij}-\delta _{ij}    
\end{equation}

\begin{figure}
    \centering
    \includegraphics[width=0.5\textwidth]{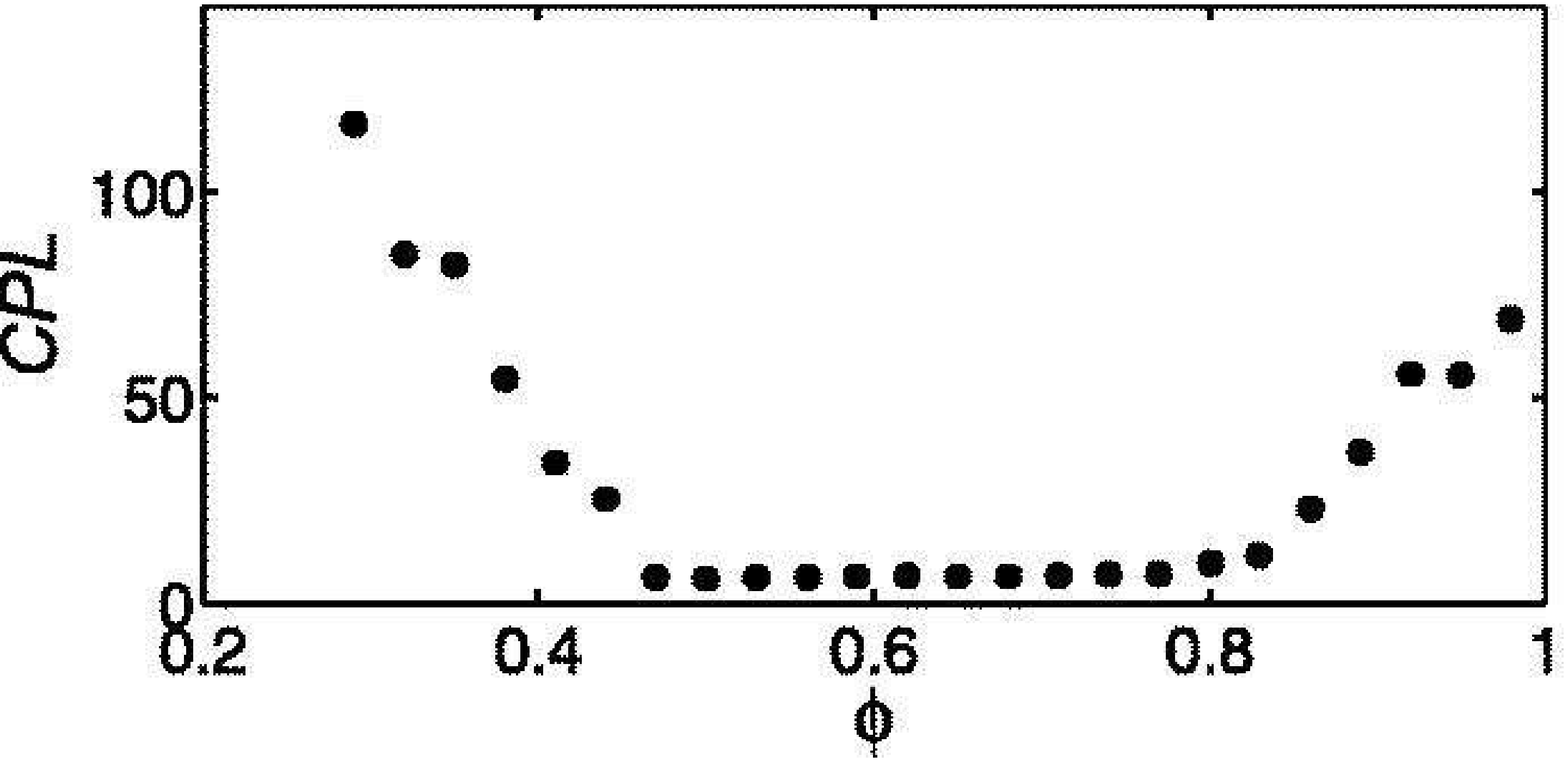}
    \caption{Variation of the CPL with the equivalence ratio, which determines the dynamical regime of the combustor. The CPL decreases as thermoacoustic instability is approached, and increases again near blowout. (Reproduced with permission from \cite{godavarthi2017recurrence}).}
    \label{fig:CPL}
\end{figure}

Then, in general, all the known complex network measures, given in Table \ref{table:network_measures}, can be used to indicate transitions, and they can support and complement the measures from RP \cite{marwan2013recurrence, ambika2020methods}. We note that the changes in dynamics change the pattern of recurrences of the trajectory points in the state space and such changes will reflect as changes in values of RN measures. Godavarthi et al. \cite{godavarthi2017recurrence} have identified different dynamical regimes with difference in topologies of the associated RNs for a thermoacoustic combustor, and showed that dynamical transitions are reflected as changes in the network measures such as characteristic path length, CPL (see figure \ref{fig:CPL}). Here, CPL is less for periodic dynamics and increases due to chaotic nature or stochasticity in the dynamics. Clearly, the trends in network measures depend on the direction of transition, and prior knowledge of the system dynamics will be very useful.

\begin{figure}
    \centering
    \includegraphics[width=0.7\textwidth]{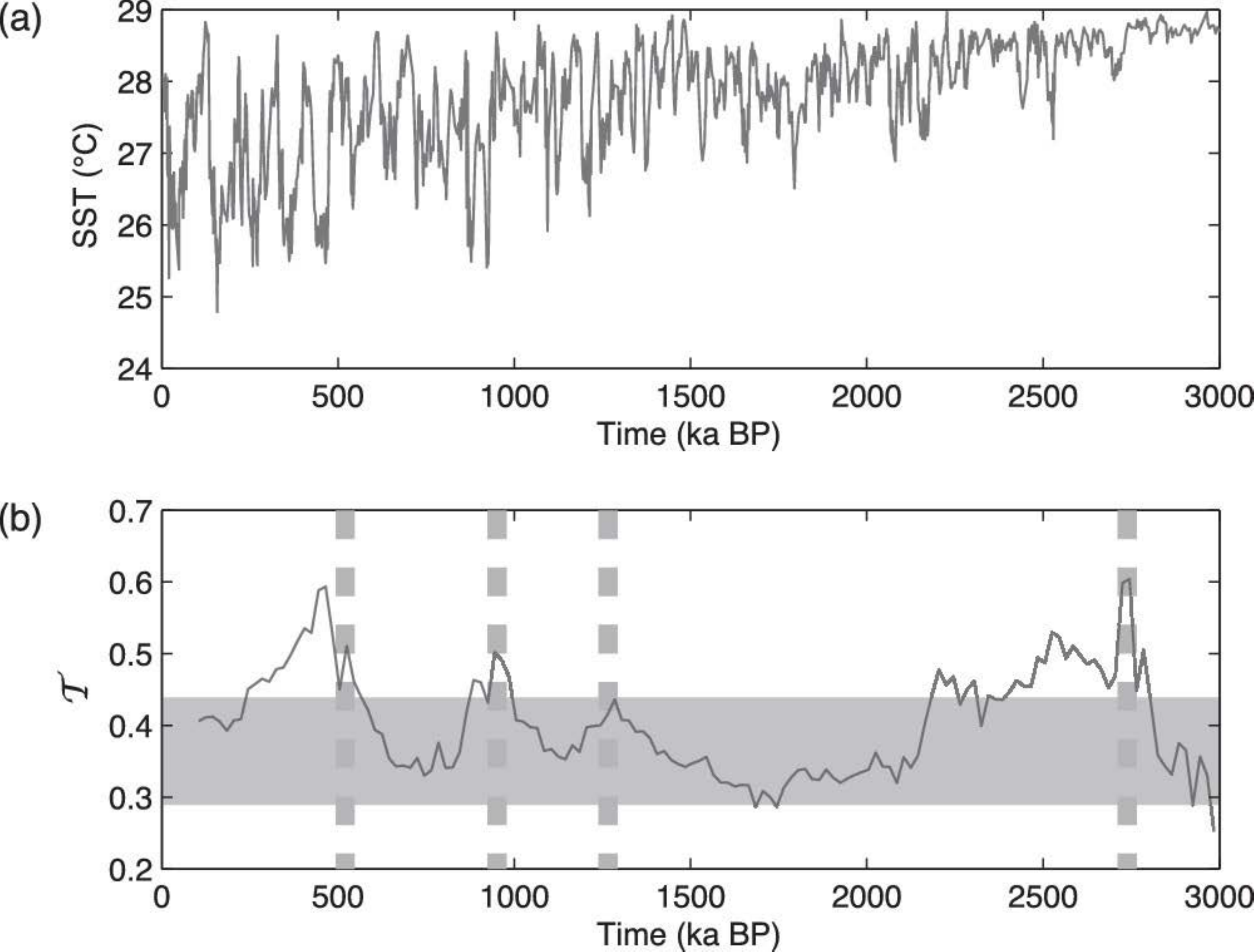}
    \caption{Detecting sudden transitions using Recurrence Network measures. (a) Alkenone paleothermometry data used as proxy for Sea surface temperature (SST). (b) Sliding window analysis of Transitivity(T), showing significant changes near major climate shifts (indicated by dashed lines). The grey band represents confidence interval of 90\% (Reproduced with permission from \cite{marwan2015complex}).}
    \label{fig:RN_south_china}
\end{figure}

Donges et al. \cite{donges2015non} have used a combination of network measures on RN to identify nonlinear regime shifts in Asian monsoon. Goswami et al. \cite{goswami2018abrupt} have proposed community structure in recurrence networks as a way to detect abrupt transitions in data with uncertainties, such as global stock indices and El Ni\~{n}o-Southern Oscillation. An interesting study by Marwan et al. \cite{marwan2015complex} shows a significant increase in network transitivity leading to sudden transition in the alkenone paleothermometry data (see Figure \ref{fig:RN_south_china}). 

The notion of recurrence networks can be extended to multivariate data as well, as illustrated by Hasselman et al. \cite{hasselman2022early}. They used Multiplex Cumulative Recurrence Networks as model networks that capture changes in mental state for the same data set as discussed in \cite{wichers2016critical} (see figure \ref{fig:depression}). The state of the individual is captured across five variables, namely Mood (Mo), Physical (Ph), Self Esteem (SE), Mental Unrest (MU), Sleep (Sl) and Day (Da), and recurrence is calculated cumulatively across all past time points. cRNs constructed this way for all five variables are then multiplexed in a single Multiplex Recurrence Network (MRN). The topological properties of this MRN show indications of transition prior to the depression episode. \\

\section{Prediction of transitions using Machine Learning methods}
\label{sec:ml}
In recent years, with increase in the size and number of available data sets, it has become computationally efficient to invoke machine learning algorithms to predict critical transitions. Unlike physics based methods, such as CSD, these methods are purely data-driven and are often difficult to interpret. However, they have shown remarkable success in predicting sudden transitions, and are well suited to deal with higher-dimensional data. Much of the work that deals with predicting critical transitions is fairly recent, and mostly focuses on the ability of the method to do so in simulated dynamical systems.

While a detailed discussion of machine learning is beyond the scope of this review, we briefly introduce the terminology and methods that will be mentioned in this section. We start from the data to be analysed, which is represented using a number of its features in an n-dimensional feature space. In supervised learning, which forms most of the cases discussed in this section, this data is split into a training and a testing set. A machine learning algorithm learns on the training data by varying model parameters such that a suitable loss function, that measures how different the true output is from the estimated output, is minimised. The model performance can be estimated by quantifying how well the model predicts the unseen testing set. 

These models differ from each other in terms of how they fit training data. A linear regression model for instance would fit a line to the training data, such that the error is minimised. The fitted line can be used to predict the output for an unknown dataset. Support vector machines classify data in a high dimensional feature space by drawing a boundary that maximises the distance of the categories from the boundary. A decision tree uses data features to learn simple decision rules that are used to predict target values. A random forest is an ensemble of decision trees trained on subsets of features and data to classify data. The class predicted by the most number of trees is chosen as the class predicted by the forest. Artificial neural networks consist of layers of connected neurons whose weights are trained to obtain maximum prediction accuracy. Numerous variations and improvements which are adapted for working with various types of data exist for each of these models. For instance logistic regression models can handle data with binary outcomes, convolutional and recurrent neural networks work on image and time series data respectively and gradient boosted trees improve on conventional random forests.

Many of these techniques including decision trees and random forests \cite{hyland2020early, tapak2019comparative}, support vector machines\cite{kobayashi2019early}, convolutional neural networks(CNN) \cite{bury2021deep, lapeyrolerie2021teaching, deb2022machine}, and recurrent neural networks (RNN) \cite{guo2020early, kong2021machine} are used to predict sudden transitions. For instance in Hyland et. al., the task was to predict circulatory failure from ICU data\cite{hyland2020early}. A set of features involving static features such as the age or sex, history of instability, summaries at different time resolutions and shape let based features were used to train a number of machine learning algorithms, including gradient boosting trees, logistic regression and RNNs. In Kobayashi et. al., a feature set derived from ordinal partition transition networks was used to detect combustion instability using support vector machines\cite{kobayashi2019early}. 

\begin{figure}
    \centering
    \includegraphics[width=0.75\textwidth]{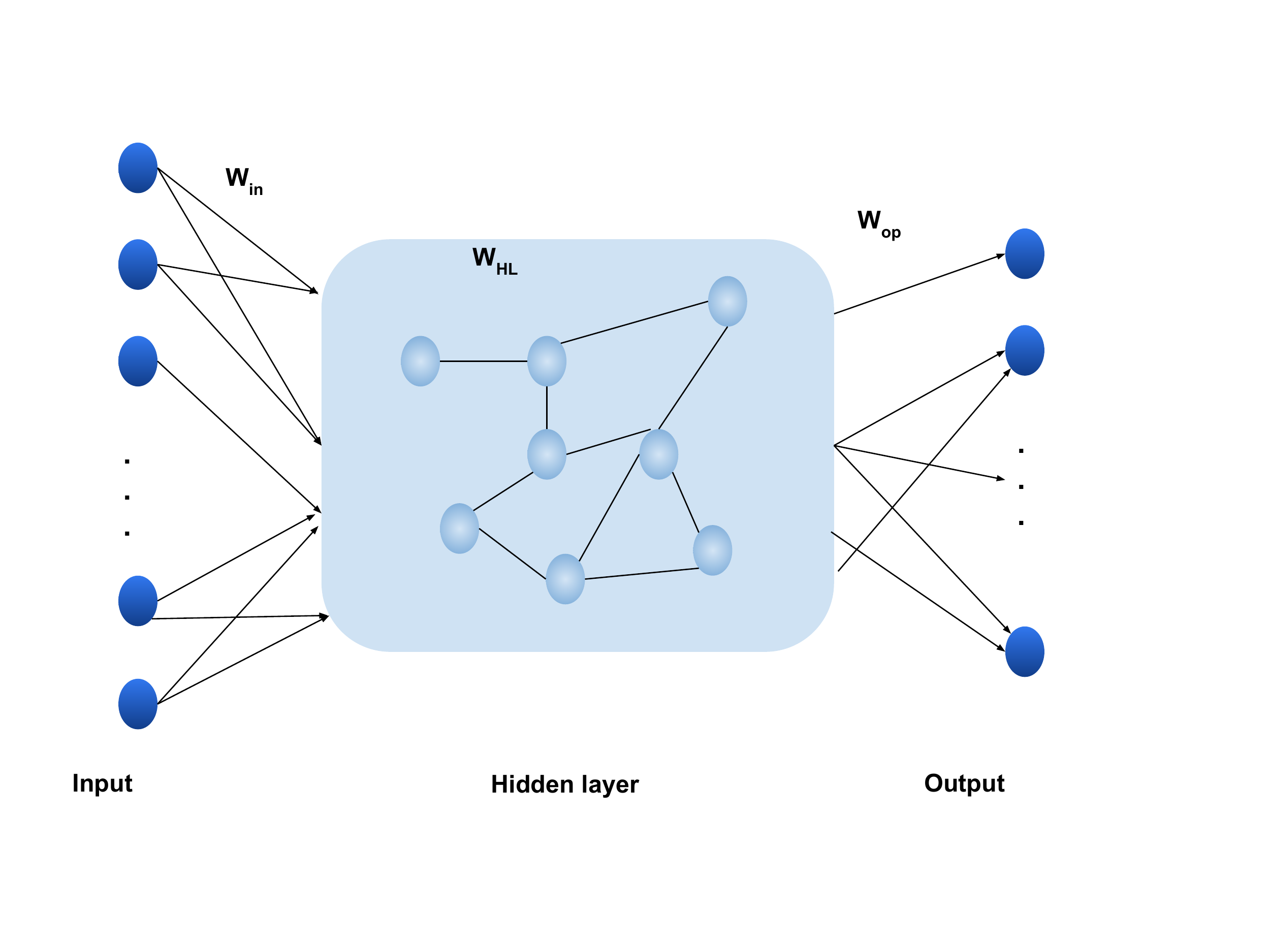}
    \caption{Schematic of a reservoir computer depicting the input, hidden and output layers. Unlike conventional neural network architecture, only the weights of the output layer are trained in reservoir computers. }
    \label{fig:ESN}
\end{figure}

As opposed to approaches that use features derived from data, RNNs learn patterns in time series themselves to predict and classify them. So they are among the most successful machine learning approaches to predict sudden changes in time series. RNNs typically possess a cyclic connection that makes the current state dependent on past states. A number of RNN architectures exist, such as discrete-time recurrent multilayer perceptrons, gated recurrent units, long short term memory (LSTM) and echo state networks (ESN)  \cite{tsoi1997recurrent, yu2019review}. Discrete-time recurrent multilayer perceptrons are used to predict epileptic seizures \cite{petrosian2000recurrent}, gated recurrent units are used to predict heart failure onset \cite{choi2017using} and financial crises \cite{tolo2020predicting}, and LSTM networks are used to predict heart failure \cite{hyland2020early}, financial crises \cite{tolo2020predicting}, and rock bursts \cite{di2021rock}. Also, combining RNNs with CNNs is shown to be effective in predicting instabilities when the data is spatio-temporal in nature \cite{gangopadhyay2020deep, lyu2021comprehensive}. This approach is useful, because convolutional neural networks are well suited to learn patterns in spatial data, whereas recurrent neural networks are well suited to time series. 

Reservoir computers or echo state networks (ESN) are a class of RNN that have shown particular promise in the prediction and classification of nonlinear dynamical systems \cite{jaeger2001echo}. These consist of an input layer, a hidden layer with random connections and an output layer. In contrast to a regular RNN, the input and hidden layer weights are not modified during training, and only weights of the output layer are trained (See Figure \ref{fig:ESN}). Recent work by Kong et. al \cite{kong2021machine} used a modified reservoir computer to predict critical transitions from a chaotic state to collapse via crisis transitions in simulated dynamical systems. In addition to the regular input of the state of a dynamical system, an additional input channel was added for the parameter of the system. This made the reservoir computer ``parameter aware", and trained the algorithm on data as well as the parameter associated with that data. Such a modified reservoir computer was able not only to predict the point where the critical transition will occur, but also to predict the distribution of the lifetimes of the chaotic transients in the parameter regime beyond the transition point. Such `parameter-aware' reservoir computers are shown to be able to go beyond prediction of chaotic transients and collapse by Patel et. al \cite{patel2021using}, where the overall dynamics of different chaotic systems over a range of parameter values, were predicted with a high degree of confidence. The authors were also able to modify the input and prediction of the reservoir computer to accommodate noisy chaotic systems. The parameter-aware reservoir computers are also reported to predict the synchronisation and amplitude death in dynamical systems \cite{fan2021anticipating, xiao2021predicting}. In the former, the system was able to predict the transition point for both smooth and explosive synchronisation transitions by training the reservoir computer on a group of asynchronous time series before the onset of synchronisation\cite{fan2021anticipating}. In the latter, the parameter-aware reservoir computer was trained on data prior to amplitude death, and was able to predict the onset of amplitude death in a number of simulated systems\cite{xiao2021predicting}. Moreover, the predictability, defined as the ability of the ESN to predict the state of the system at a point in future, is also shown to decrease leading up to critical transitions\cite{choi2022early}. Deep extensions of reservoir computers, where the reservoir consists of stacked layers where each layer feeds into the next, are studied to predict critical transitions in simulated slow-fast systems\cite{lim2020predicting}. 

Machine learning was also used recently to predict critical transitions in complex networks\cite{ni2019machine,grassia2021machine}. Feed forward neural networks (neural networks without a cyclic connection) are used to study critical transitions in models of epidemic spreading in complex networks. The model trained on the states of the node dynamics, with labels indicating  whether the network was before or beyond the critical point. The method could not only anticipate a phase transition, but also identify the critical point itself\cite{ni2019machine}. A graph attention network with a feed forward regressor is shown to be useful in predicting disintegration in complex networks. The network was trained on the topological structure of different synthetic and real world complex networks, and could anticipate system collapse\cite{grassia2021machine}.

\begin{figure}
    \centering
    \includegraphics[width=0.75\textwidth]{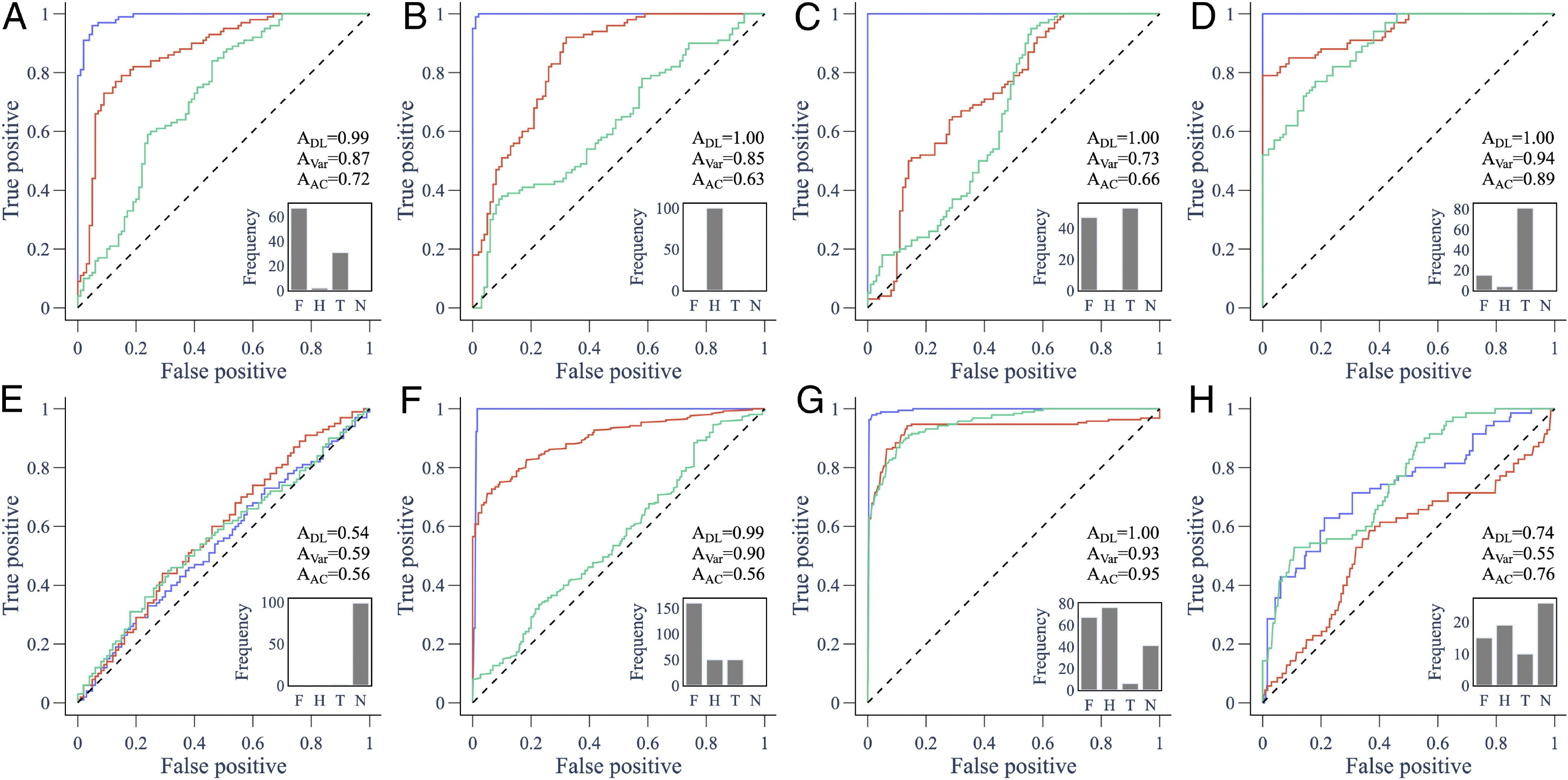}
    \caption{Prediction accuracy quantified using the area under the ROC curve for ACF(1), variance and a deep learning model. The inset shows the probability predicted for a fold (F), transcritical (T), Hopf (H) or no (N) bifurcation, by the model. The data are from (A) fold bifurcation in a harvesting model; (B ) Hopf and (C) transcritical bifurcation in a consumer-resource model; transcritical bifurcations from two variables, (D) provaccine opinion and (E) total infectious from a behavior-disease model; (F) sediment data transitioning to anoxic states in the Mediterranean sea; (G) Hopf bifurcation in data from a thermoacoustic system; and (H) transitions in ice core records. (Reproduced with permission from \cite{bury2021deep})}
    \label{fig:EWS_ML}
\end{figure}

As mentioned earlier, while machine learning techniques are being used extensively for predicting critical transitions, they are model-free and do not explicitly work on the principles of CSD. Recently, some approaches have been adopted which combines principles of CSD with machine learning to predict critical transitions\cite{ma2018data, lassetter2021using, bury2021deep, deb2022machine}. In \cite{ma2018data}, three EWS measures, namely the variance, autocorrelation and skewness are used as input features for a support vector machine, while the bifurcation point is used as the output. The trained algorithm showed high accuracy when predicting the bifurcation point in multiple realisations of a simulated stochastic slow-fast dynamical system, as well as in the IEEE 14-bus test system. In \cite{lassetter2021using} the accuracy of machine learning models enhanced with EWS was compared to a model without any inputs about the EWS. Using EWS data to train the neural networks was shown to enhance the performance in most of the cases. In \cite{bury2021deep} and \cite{deb2022machine}, the authors  compared the performance of a well trained convolutional neural network-long short-term memory (CNN-LSTM) deep learning algorithm with conventional EWS such as the auto-correlation and variance in predicting critical transitions. In \cite{bury2021deep}, the neural network was trained using data from randomly generated 2-D dynamical systems containing polynomial non-linear terms up to cubic order. Then the bifurcations in these dynamical systems were identified and the time series were generated. The results suggested that the neural network performed well in predicting critical transitions arising from different bifurcations in both simulated and real data \cite{bury2021deep}. The prediction accuracy for standard EWS and the machine learning model for various simulated and real-world data are shown in Figure \ref{fig:EWS_ML}. The machine learning probabilities calculated from raw data assigns the three transitions correctly in almost all cases. In \cite{deb2022machine}, nine different dynamical models from a variety of fields were used to train a CNN-LSTM model, called EWSNet. The trained network was then used to identify transitions in simulated as well as real-world data, achieving high classification accuracy in both. In addition, the authors compared the results to various machine learning classification algorithms trained on trends in classical early warning indicators and demonstrated that the CNN-LSTM architecture showed superior performance. A python implementation of EWSNet may be found on https://ewsnet.github.io/.

In contrast to equation based models and real world time series that are generally used to study EWS, \cite{fullsack2020training} studied warning signals in an agent based model, using an LSTM neural network architecture. The authors used large datasets of time series simulated using an agent based model of repeated public good games, which showed critical transitions. These were used to train an LSTM network, and predict transitions in an unseen simulated dataset. The prediction was studied as the distance to the transition was reduced, and the results were compared to traditional EWS metrics. The authors also studied the micro level interactions using time series from each node of the agent based model.

Machine learning algorithms have added an important tool in the prediction of critical transitions. Unlike critical slowing down, these are model independent, and can hence tackle a wider range of transitions and even extreme events\cite{ray2021optimized, meiyazhagan2021model}. However, they require large datasets for training, that is often not possible in real world scenarios. Powerful new techniques that combine differential equation models along with machine learning is a vital middle path that has shown promise in recent years, and could be the way forward in efficiently predicting critical transitions.

\section{Reliability of measures as EWS}
\label{sec:reliability}
 
As mentioned, much of EWS analysis is conducted using a sliding window approach over sufficiently long data sets.
However, this method is prone to misdetections arising from data quality, as well as from the inherent dynamics of the system. For instance, systems with a longer response time require more data before a transition in order to identify warning signals. This means that the amount of data required seems to depend on the total range of system dynamics captured and not on the length of the data alone \cite{van2021no}. Even with the same length of data, the number of points required to make a reliable estimation of a quantifier varies between the different quantifiers being used (refer Table \ref{table:error}). \\

The effects such as the topology of connected systems, under-sampling or insufficient data before the transition \cite{jager2017hidden, wen2018one}, can lead to false negatives. False positives on the other hand, can occur due to a number of reasons, such as serial correlations due to overlap in the windows or due to the presence of trends in data \cite{hamed1998modified,jager2019systematically}. The effect of serial correlations in data when testing for significance using the Kendall $\tau$ correlation was studied by Hamed and Rao \cite{hamed1998modified}. Systematic false positives occurring due to trends in the data can sometimes be removed by using detrending techniques such as local first-order polynomial fit, global higher-order polynomial fit or moving averages \cite{jager2019systematically}. Conversely, the features introduced by particular types of detrending techniques may be required by deep learning algorithms, in order to effectively detect critical transitions \cite{dablander2022deep}. A final possibility is when the transition does not occur due to various reasons, even when CSD did occur, in reality. For instance, warning signals may be correctly predicting seizure susceptibility, but no seizure may happen due to the internal regulatory mechanisms of the brain \cite{freestone2017forward}. These are not false positives in the true sense but will be treated as such in many studies. \\
 
 An alternative approach to test for significance, with lesser misdetections is bootstrapping or surrogate testing \cite{small2003detecting}. First, multiple surrogate datasets consistent with a null hypothesis is generated based on a property of the original dataset. The commonly used surrogates include randomly shuffled surrogates, autoregressive model-based ($AR(1)$) surrogates and iterated amplitude adjusted Fourier transform (IAAFT) surrogates \cite{dakos2012methods}. Randomly shuffled surrogates preserve the amplitude distribution of the data, and hence the global moments of the time series. AR(1) surrogates generate multiple instances of an AR(1) process, preserving the ACF(1) and the variance of the original dataset. The IAAFT surrogates preserve the frequencies present globally in the original data while randomising the Fourier phases in the data. The significance is tested based on the percentile of the correlation coefficient of the original dataset in comparison to the distribution of correlation coefficients in the surrogates. The bootstrapping method was successfully implemented with recurrence based EWS for significance testing for climate transitions from marine proxy record of sea surface temperature \cite{marwan2013recurrence}.

Conventional EWS are resilience indicators, such as recovery rates from small perturbations. Sometimes this resilience is confused with the \textit{stability} of a system state, which actually corresponds to the width of the basin of attraction for that state \cite{dakos2022ecological}. In this context, Nes et al. \cite{van2007slow} suggest that the relationship between the width of the basin and the recovery rate has a specific slope for a given system, depending strongly on the timescales of its components. Hence, situations such as rapid changes in environment \cite{dai2015relation}, or large external influences may not show the correlation between the two. Moreover, there are cases where the system parameters change too rapidly for the system to settle. In these cases, detection through EWS becomes a question of speed as well as reliability.\\

\section{Outlook and Perspective}
\label{sec:dis}
In this review, we present an overview of the set of measures that can function as warning signals in predicting critical transitions in complex systems. Apart from conventional EWS that arise from CSD, we also discuss recent trends in detecting critical transitions in extended and connected systems. In spatial systems, these are often studied using spatial extensions of commonly used EWS such as spatial autocorrelation and variance. We anticipate and encourage more applications using insights from SEWS. There are many situations where spatial data is relevant, although not at a scale as big as ecosystems. Thus neuroscience, transportation systems, electric power grid etc. are some of the areas where spatial order matters, and critical transitions are highly relevant. For general extended systems, methods based on multivariate quantifiers and complex network measures are also discussed. 

In general, conventional measures perform well when the transitions are preceded by CSD. However in cases of noise-induced or rate-induced transitions, the measures based on patterns of recurrences offer a new set of measures that are effective for short and non-stationary data sets \cite{ioana2014recent}. We find that trends in DET, LAM and CPL can give indications of these transitions that can be confirmed with supporting evidence from conventional measures, when applicable. The utility of recurrence-based measures comes from their ability to detect the nature of dynamical regimes that the system displays, indicating the type of transition. 

A major thrust in the recent past, has been to use machine learning and deep learning methods to study and predict critical transitions. These methods have shown considerable promise in learning and predicting the dynamics of nonlinear systems, including predicting upcoming critical transitions. In this context, we included details on how techniques from machine learning, such as reservoir computers and convolutional neural networks are used to predict upcoming critical transitions. 

However, an important aspect to be considered while applying the above mentioned EWS is the reliability of detected EWS as indicators of critical transitions, and how to decrease misdetections. We discussed how to avoid false negatives and be careful about false positives. We now proceed to describe some recent under-explored developments in the field of EWS, and some areas that hold scope for further research in the field.

In biological systems, a separate line of studies exist focusing on dynamical network markers. Here, critical transitions are associated with complex diseases such as lung injury, cancer \cite{chen2012detecting}, and type-1 diabetes \cite{liu2013detecting}, where a pre-disease state can be identified using dynamical network biomarkers (DNBs). A rapidly emerging area is infectious disease spread, fueled by the urgency of issues rising from COVID-19 pandemic \cite{southall2021early}. Some other areas where EWS framework has shown promise are chemical reaction networks \cite{maguire2020early}, biological systems \cite{clements2018indicators}, gene expression dynamics\cite{pal2013early}, and socio-ecological systems \cite{suweis2014early}. The development of new EWS in these and related areas, as well as adapting existing frameworks to them, should open new exiting doors with applications to real complex systems and processes.

In recent studies, by assigning strengths to connections based on cross-correlations, global teleconnections are derived from global climatological data like sea surface temperature, that predict and uncover rainfall anomalies, draughts, disease outbreaks, El Ni\~{n}o etc. \cite{boers2019complex, agarwal2019network, anyamba2012climate}. However, due to the large range of spatial and temporal scales involved, this area still poses many challenges. The construction of multilayer or multiplex networks from multivariate or spatial data can be a good approach here, and in similar contexts, that can provide better and useful predictions.

Many complex systems have active tipping points of interacting multi stable systems modelled by complex networks. Then the topological features of the network also play a role in inducing transitions and hence are useful in deriving measures of EWS. In this context, the vulnerability of the tipping networks to cascades depends on the topology and hence is important to consider features like clustering that can trigger cascades. Identifying generic indicators based on complex networks will be very interesting since modelling complex systems in a quantitatively accurate way is still difficult to achieve. Moreover, whether a typical node of the network will start tipping first, followed by others or is it an integrated cooperative transition affecting all nodes at the same time is not always very clear. Hence for predicting instability and vulnerability in such complex systems, several  targeted research efforts are required.

There are a few instances where modelling complex systems using dynamical systems on complex networks can give relevant information and can model dynamics of real-world systems. Recent studies establish sudden transitions and tipping in such systems, like explosive synchronisation and explosive death, multiplexing induced critical transitions etc. \cite{zhang2014explosive, dixit2021dynamic, verma2021tipping}. In these theoretical models of coupled dynamical systems, explosive synchronisation is associated with hysteresis \cite{d2019explosive}. The most general pathway to explosive synchronisation is adiabatic change in the coupling parameter. Such a transition is peculiar because there is little to no indication when the coupling parameter approaches the threshold of synchrony, and then suddenly the whole network synchronises. While the order parameters and other characteristics in such transitions are well studied, possible EWS for these transitions remain elusive. 

In the context of purely model based studies of EWS, agent-based models are suggested as complementary to equation-based models \cite{Reisinger2020Comparing}. Using this approach, F\"{u}llsack et al. have highlighted subtleties in critical transitions arising from complex interactions obscured in equation-based models \cite{fullsack2022early}. A very recent line of work explores hidden transitions in multiplex networks, that carry signatures of first-order transitions \cite{da2022hidden}. It would be interesting to identify EWS in this scenario, with possible applications to multivariate data.

Understanding the underlying dynamical processes that induce tipping is very important both for a research point of view and for applications in the design of proper interventions and later for monitoring the recovery process. While data-driven studies cannot be of much use, the dynamical system approach based on standard systems or models can throw light on them. Arriving at models will also help to isolate the underlying processes that generate the critical transitions. However, historical data can help immensely in establishing `normal' patterns of behaviour in natural systems that are considered to be `stable' for a long time. In this case, data-driven approaches can help generate better suited models of such systems in which parameters can be varied artificially and impact on the system can be studied in simulations. Climate change,  socio-economic developments and finance are the thrust areas for such modelling approaches. 

Even though the trends in various computed measures are interpreted as an indication of upcoming critical transitions, the details of the trends and their variations are not studied. They can throw light on the mechanisms underlying the transitions and can give a finer level of characterisation. One of the perspectives in this direction can be isolating the scaling behaviour of individual EWS measures near the transitions and the possibility of their use in identifying the nature of transitions based on the corresponding scaling indices. Another approach in this direction is suggested by Halekotte et al.\cite{halekotte2020minimal} focusing on minimal fatal shock, \textit{miFaS} (minimal perturbation) that can change the state of a system suddenly. This \textit{shock-induced tipping} quantifies the magnitude of the shock as a global stability measure that can provide information on weak points constituted by certain substructures of the network and to understand their topological and dynamical origin.

The availability of multiple EWS allows for cross-checking and parallel analysis that can prepare us for anticipated sudden changes and hence reduce the risks and irreversible and irreparable damages. In a few special cases, this also encompasses chances for promoting desired transitions. However, the time frame involved in arriving at EWS is very crucial in many contexts for public awareness and adaptive or preventive interventions. More studies are needed to develop EWS that provide information on the time frame and/or the probability of the transition.
Such an EWS only can lead to specialised mitigation strategies \cite{biggs2009turning}. This would require choice of suitable thresholds for any of the measures beyond which tipping can be confirmed in the system. With an increase in the size and number of data sets to be analysed, this has become a real challenge. In this direction, various machine learning approaches are increasingly becoming useful to extract common features. This has an additional advantage of computing EWS in real-time fast enough to be useful for starting mitigation strategies and related procedures well in advance.

Also more attention needs to be on the transitions that have no indication of CSD but can be stochasticity driven \cite{diks2019critical}. In financial markets, contradicting evidence exists for CSD prior to sudden transitions \cite{nawrocki2014bifurcation, kozlowska2016dynamic,diks2019critical}. Very recently, dynamics of currency and crypto-currency exchange rates has caught on with studies of EWS \cite{wen2018one, tu2020critical}, and evidence of transitions without CSD \cite{guttal2016lack}.
Recent studies indicate the occurrence of such transitions in the financial crisis, where an early prediction is highly useful and sought after \cite{harikrishnan2022recurrence}. The nature of the processes that lead to this type of transitions and mechanisms underlying them require detailed further studies.  

A related objective will be a good estimate of risk which is the product of the probability of transition and its negative impact or losses. While studies on EWS that can predict the transition have advanced in many cases, the ignorance on impacts persists. This will help to design protocols to reduce risks. Some of the other challenges for the future would be to identify and reduce chances of false positives and false negatives. A possible future direction is traced by Laitinen et al., taking probabilistic approach to EWS detection \cite{laitinen2021probabilistic}.

With the literature in critical transitions and EWS growing, it is important to understand the overall context of some of these concepts and their impact in combating daunting issues such as climate change and socio-ecological systems that involve challenges present at different spatial and organisational scales \cite{lenton2022resilience}. These issues need immediate attention in terms of development of estimation of risk and mitigation strategies. Fortunately, computational resources and widespread availability of data make it possible to address them. A unified framework incorporating data-based and analytic approaches, like synergy of data science, network science and EWS theory, is crucial for arriving at new measures from data. So also, optimising and designing automated warning systems, machine learning based detection, testing of EWS with large data sets to judge their practical applicability and unforeseen issues are relevant for further research.

\section*{Acknowledgements}
The authors would like to acknowledge the useful discussions at the satellite session on ``Early warning signals in complex systems" organised by them at the Conference on Complex Systems 2020.
SVG thanks the TRANS-ID team at the UMCG for useful discussions on critical transitions and early warning signals.
\section*{References}
\providecommand{\newblock}{}

\end{document}